\documentclass[11pt]{elsarticle}
\usepackage{graphicx,color,amsmath,amssymb}
\usepackage{bm}
\usepackage[colorlinks=true, pdfstartview=FitV, linkcolor=red, citecolor=blue, urlcolor=blue]{hyperref}
\usepackage{ulem}

\usepackage{lipsum}

\makeatletter
\def\ps@pprintTitle{%
 \let\@oddhead\@empty
 \let\@evenhead\@empty
 \def\@oddfoot{}%
 \let\@evenfoot\@oddfoot}
\makeatother

\usepackage{amsfonts,amsbsy}
\usepackage{amssymb}
\usepackage{graphicx}
\usepackage{color}
\usepackage{amsmath}

\def\itpile#1\over#2{\mathrel{\mathop{\kern 0pt#1}\limits_{#2}}}

\def\beq{\begin{equation}}
\def\eeq{\end{equation}}
\def\bea{\begin{eqnarray}}
\def\eea{\end{eqnarray}}

\newcommand{\Lb}{\left(}
\newcommand{\Rb}{\right)}
\def\p{{\boldsymbol p}}

\def\x{{\boldsymbol x}}

\def\d3p{\frac{d^3\p}{(2\pi)^3}E_\p}

\begin{document}

\begin{frontmatter}
\title{The {\it m}'bottom-up parton system with two momentum scales}


\author[label1]{Vladimir Khachatryan}
\author[label2]{Mickey Chiu}
\author[label1]{Thomas K. Hemmick}

\address[label1]{Department of Physics and Astronomy, Stony Brook University, Stony Brook, NY 11794, USA} 
\address[label2]{Physics Department, Bldg.\,510A, Brookhaven National Laboratory, Upton, NY 11973, USA}

\begin{abstract}
One possible evolutionary scenario of the dense gluon system produced in an ultrarelativistic heavy ion collision
is the bottom-up thermalization scenario, which describes the dynamics of the system shortly after the collision
via the decay of originally produced hard gluons to soft ones through QCD branching processes. The soft gluons
form a thermal bath that subsequently reaches thermalization and/or equilibration. There is a scaling solution
to the bottom-up problem that interpolates between its early stage, which has a highly anisotropic gluon distribution,
and its final stage of equilibration which occurs later. Such a solution depends on a single parameter, the so
called momentum asymmetry parameter $\delta$. With this scaling solution, the bottom-up scenario gets modified
and the evolving parton system, referred to as the {\it m}'bottom-up parton system throughout this paper, is
described by this modification. The time evolution of the system in the original bottom-up ansatz is driven by
the saturation scale, $Q_{s}$. However, for the {\it m}'bottom-up we generalize the ansatz of the evolution
by introducing two additional momentum scales, which give a thermalization time and temperature of the soft gluon
bath somewhat different from those obtained when the {\it m}'bottom-up matches onto the final stage of the 
original bottom-up scenario.
\end{abstract}


\end{frontmatter}

\section{Introduction}

The original bottom-up thermalization ansatz \cite{Baier:2001} in its basis has the parton saturation mechanism expected to be valid at high parton densities. It leads to thermalization/equilibration of a parton system produced after a heavy ion collision. This scenario emphasizes the importance of branching processes of initially produced gluons, and as a consequence of such processes the total number of gluons increases between the initial and thermalization times of the evolution of the system.

The bottom-up does a good job in predictions of charged hadron multiplicities observed at RHIC and LHC energies \cite{Baier:2002}. However, it should be noted that the overall bottom-up picture can also be modified \cite{Mueller:2006,Mueller:2006II} to include other possibilities for evolution of the produced matter, which we shall refer to as {\it modified} bottom-up parton system or simply as {\it m}'bottom-up parton system. The modification is done by a scaling solution to pre-equilibrium evolution which interpolates between initial plasma instabilities \cite{Arnold:2003,Mrowczynski:2007,Mrowczynski:2006} present in the dense gluon system produced immediately after a collision, and the final equilibration. Depending on a single parameter known as the momentum asymmetry parameter $\delta$, the scaling solution matches onto the original bottom-up picture either at an intermediate stage or toward the end of the evolution given by the original bottom-up.

Thus, by having the original picture modified we further argue that in the {\it m}'bottom-up scenario the dominant qualitative and semi-quantitative features of the evolution of a parton system can be described by two momentum scales, $\Omega$ and $\Omega_{s}$. Note that in the original bottom-up the evolution is described only by one scale - the saturation momentum $Q_{s}$. The situation somewhat resembles the overpopulated Glasma, where two dynamical scales $\Lambda$ and $\Lambda_{s}$ are introduced to describe the dynamics of the system from the overpopulated initial stage all the way to thermalization \cite{Blaizot:2011xf}. In this paper we show that the scales $\Omega$ and $\Omega_{s}$ can describe the dynamics of the {\it m}'bottom-up parton system from its initial stage toward thermalization. However, depending on the parameter $\delta$, the thermalization time and the thermalization temperature obtained, can be of the same order (or different) as compared with those derived in the original bottom-up scenario. The results of this paper can be useful for calculating the yield of direct photons and di-electrons as well as the elliptic flow of direct photons, in particular, in Au+Au collisions at $\sqrt{s_{NN}} = 200\,GeV$ collision energy, measured by the PHENIX experimental collaboration \cite{Adare:2008fqa,Adare:2009qk,Adare:2011zr,Adler:2003}. 

In the next section we give a brief overview of the Glasma dynamical scales. In the third section we derive some quantities, which describe quark-gluon matter, by using the dynamical scales $\Omega$ and $\Omega_{s}$ of the {\it m}'bottom-up scenario. Besides, we also show that these and Glasma scales are the same at least on the parametric level. In the last section we show some estimates of the thermalization time and temperature at 200\,GeV.

\section{The two momentum scales of the Glasma}

First, let us discuss what happens in the Glasma. The Glasma is theorized to exist from the earliest time after a heavy ion collision when the fields are considered to be highly coherent, and when most of the energy is in coherent field degrees of freedom rather than in the degrees of freedom of incoherent quarks and gluons. Thereby, the matter in this phase is neither the Quark-Gluon Plasma nor the Color Glass Condensate, however, it has features of both. Almost instantaneously after the collision, the transverse color fields of the Color Glass Condensate transform into the longitudinal color electric and color magnetic fields of the Glasma \cite{Lappi:2006fp}. In this scenario the gluons, in the sense of particles, are produced from the classical evolution of such color electric and magnetic fields, and after initial plasma instabilities produce a distribution that is approximately isotropic in momentum space. 

The evolutionary and thermalization processes of the Glasma and {\it m}'bottom-up scenarios are different from each other because of the underlying mechanisms by which they work. Nonetheless, they can have some properties which are similar phenomenologically, and we shall investigate these properties in this paper. In this section we follow the lines of reasoning represented in Refs.\,\cite{Blaizot:2011xf} and \cite{Chiu:2012}.

It is expected that on the transverse momentum and mass scales the effects stemming from evolution to the thermalized distribution of gluons are enhanced. The evolution of the system toward thermalization can be traced based on the Boltzmann transport equation:
\begin{equation}
\partial_{t}f(\textbf{\it{k}},x) = C_{\textbf{\it{k}}}[f]\,,
\label{eqn_Boltzmann}
\end{equation}
where the $C_{\textbf{\it{k}}}[f]$ is the collision integral, and the $f(\textbf{\it{k}},x)$ is the particle distribution function. The main qualitative features of the solution of the Boltzmann equation can be described if it is assumed that the evolution is dominated by only two scales - the ``infrared" $\Lambda_{s}$ and  ``ultraviolet" $\Lambda$, characterizing the gluon distributions in the Glasma. It is also assumed that the elastic collisions play a dominant role in driving the initial gluon distribution toward local equilibrium. 

The scale $\Lambda_s$ is a momentum scale at which the distributions are highly coherent, and it is time dependent. The scale $\Lambda$  above which the distributions become dilute is also time dependent, which coincides at the earliest time with $\Lambda_{s}$: $\Lambda(\tau_{0}) = \Lambda_{s}(\tau_{0}) \sim Q_{s}$ (the $\tau$ is the proper time of the collision). Along with time these scales separate from each other where the $\Lambda_{s}$ decreases rapidly, and the $\Lambda$ evolves more slowly. Upon reaching equilibration, the $\Lambda$ becomes the initial temperature of the Quark-Gluon Plasma: $\Lambda(\tau_{therm}) \sim T_{in,QGP} $. On the other hand, the infrared scale  becomes the so called non-perturbative ``magnetic scale''  \cite{Liao:2008,Liao:2006} in the thermalized matter: \,$\Lambda_s(\tau_{therm}) \sim \alpha_{s}\,T_{therm} = \alpha_{s}\,T_{in,QGP}$,\, where the $\alpha_{s}$ is the QCD coupling constant. The thermalization is accomplished by parametrically splitting apart these initially overlapping momentum scales by $\alpha_{s}$, and the corresponding time is determined from the following requirement:
\begin{equation}
\Lambda_{s}(\tau_{therm}) \sim \alpha_{s} \Lambda(\tau_{therm})\,,
\label{eqn_Glasma_scales}
\end{equation}
at which the gluon occupation number, $f(\Lambda)$, becomes of the order of unity. Generally, the $f(\Lambda)$ is proportional to $\sim 1/\alpha_{s}$ which is down to the ultraviolet scale, whereas the thermal distribution, $1/\omega_{p}$, builds up from the infrared scale gradually. Thus, the gluon distribution becomes a thermal gluon distribution function when the infrared scale satisfies the condition in Eq.\,(\ref{eqn_Glasma_scales}).

In the momentum space there are the following cases (at time $\tau > 1/Q_{s}$) in the corresponding three ranges:
\begin{eqnarray}
\,\,\,\,\,\,\,\,\,\,& & f_{g}(p) \sim {1 \over \alpha_{s}}\,\,\,\,\,\,\,\,\,\,\mbox{at}\,\,p < \Lambda_{s}\,, 
\nonumber\\
\,\,\,\,\,\,\,\,\,\,& & f_{g}(p) \sim {1 \over \alpha_{s}}{\Lambda_{s} \over \omega_{p}}\,\,\,\,\,\,\,\,\,\,\mbox{at}\,\,\Lambda_{s} < p < \Lambda\,, 
\nonumber\\
\,\,\,\,\,\,\,\,\,\,& & f_{g}(p) \sim 0\,\,\,\,\,\,\,\,\,\,\mbox{at}\,\,p > \Lambda\,, 
\label{eqn_Glasma_distributions}
\end{eqnarray}
where the $p$ is the gluon momentum, and the $\omega_{p}$ its energy. In general, the occupation number is expressed as
\begin{equation}
f_{g} = {\Lambda_{s} \over {\alpha_{s} p}} F_{g}(p/\Lambda)\,.
\label{eqn_Glasma_occupation_g}
\end{equation}
Additionally, the gluon density and the Debye mass are expressed via the scales $\Lambda$ and $\Lambda_{s}$:
\begin{equation}
N_{g} \sim {1 \over \alpha_{s}}\Lambda^{2} \Lambda_{s}\,,
\label{eqn_Glasmagluon}
\end{equation}
\begin{equation}
M_{D}^{2} \sim \Lambda \Lambda_{s}\,.
\label{eqn_GlasmaDebye}
\end{equation}
At the initial time, by having the $\Lambda(\tau_{0})$ and $\Lambda_{s}(\tau_{0})$, these two equations are represented as
\begin{equation}
N_{g} \sim {Q_{s}^{3} \over \alpha_{s}}\,,
\label{eqn_Glasmagluon_in}
\end{equation}
\begin{equation}
M_{D}^{2} \sim Q_{s}^{2}\,.
\label{eqn_GlasmaDebye_in}
\end{equation}
At the thermalization time they will have forms represented by the initial temperature of the Quark-Gluon Plasma:
\begin{equation}
N_{g} \sim T_{in,QGP}^{3}\,,
\label{eqn_Glasmagluon_fin}
\end{equation}
\begin{equation}
M_{D}^{2} \sim \alpha_{s}T_{in,QGP}^{2}\,.
\label{eqn_GlasmaDebye_fin}
\end{equation}
The time evolution is dominated by the gluon density, and there can be some fixed asymmetry between the typical transverse and longitudinal momentum scales characterized by a parameter $\delta^{\prime}$ which is defined in terms of the longitudinal pressure:
\begin{equation}
P_{L} = \delta^{\prime}\,\epsilon\,,
\label{eqn_pressure}
\end{equation}
where \,$0 \le \delta^{\prime} \le 1/3$,\, with \,$\delta^{\prime} = 0$\, corresponding to the maximal anisotropy between the longitudinal and transverse pressure, and \,$\delta^{\prime} = 1/3$\, corresponding to the isotropic expansion. Finally, since the electromagnetic particle production arises from quark charges, the quark distribution function should also be included in the overall evolutionary picture:
\begin{equation}
f_{q} =  F_{q}(p/\Lambda)\,,
\label{eqn_Glasma_occupation_q}
\end{equation}
that brings out the quark density as
\begin{equation}
N_{q} \sim \Lambda^3\,.
\label{eqn_Glasmaquark}
\end{equation}
Initially not many quarks are present in the matter and this is suppressed by the power of $\alpha_{s}$, which is small if the $Q_{s}$ is large compared to the QCD scale $\Lambda_{QCD}$. Namely, at the earliest times, $N_{q} \sim \alpha_{s} N_{g} \ll N_{g}$,\, nevertheless, later the two densities approach each other becoming parametrically of the same order,  \,$N_{q} \sim N_{g}$. Therefore, in the time interval \,$1/Q_{s} \ll \tau \ll \tau_{therm}$\, the quark density increases to a value being of the order of the gluon density, and is no longer suppressed. Thus, the quark production at \,$\tau \sim 1/Q_{s}$\, might not be significant, or anyway not as significant as at later times.

\section{From the original bottom-up thermalization to the {\it m}'bottom-up thermalization}

\subsection{The scaling solution and the matching onto the original bottom-up}

Now let us discuss what happens in the {\it m}'bottom-up scenario. The bottom-up evolution is also based on using the solution of the Boltzmann equation which, with inclusion of the particle production, makes the parton system approach the kinetic equilibration during a time of the order of \,$\tau \sim \alpha_{s}^{-13/5}Q_{s}^{-1}$\, as obtained in \cite{Baier:2001}. The thermalization occurs in the limit of \,$Q_{s} \gg \Lambda_{QCD}$\,, which corresponds to very large nuclei or very high collision energy. Let us start with the original bottom-up, which is a mechanism that describes the system evolving toward the thermalization/equilibration after passing through three distinct stages.

\begin{itemize}
\item[] {\it Stage\,1}. Shortly after the collision, hard gluons are produced being distributed highly anisotropically. These gluons dominate in the time range of  \,$1/Q_{s} < \tau < \alpha_{s}^{-3/2}Q_{s}^{-1}$.

\item[] {\it Stage\,2}. In this phase these initial hard gluons still dominate but start emitting soft gluons via QCD branching. The soft gluons in turn equilibrate and form a thermal bath, which initially carries only a small fraction of the total energy of the system. Also, the thermal bath draws energy from the hard gluons. These processes occur in the range of \,$\alpha_{s}^{-3/2} < Q_{s}\tau < \alpha_{s}^{-5/2}$.\, The soft gluons start to overwhelm the  hard gluons, in terms of number,  at \,$\tau \sim \alpha_{s}^{-5/2}Q_{s}^{-1}$.

\item[] {\it Stage\,3}. Now the soft gluons are thermalized but continue drawing energy from the hard gluons until all the energy is removed from them and the whole system is equilibrated. The full thermalization is achieved when the primary gluons have lost all their energy to the soft gluon bath. Parametrically, this occurs at \,$\tau \sim \alpha_{s}^{-13/5}Q_{s}^{-1}$\, which is the end of the time range of this final stage: \,$\alpha_{s}^{-5/2} < Q_{s}\tau < \alpha_{s}^{-13/5}$.
\end{itemize}

However, after taking into account the existence of collective effects in the form of plasma instabilities \cite{Arnold:2003,Mrowczynski:2007,Mrowczynski:2006} in the initial stage of the bottom-up picture, one can think about a rapid thermalization scenario. It is because the plasma instabilities occurring in the dense gluon system produced immediately after a collision,  increase the amount of energy transformation from initially produced hard modes to soft modes radiated by the hard ones. Meanwhile, it has been shown in \cite{Arnold:2004} that the full equilibration time in the presence of the instabilities is not much shorter relative to that of the bottom-up. But these instabilities cannot lead directly to the equilibration since they would give an equilibration time parametrically of the order of \,$Q_{s}\tau \sim 1$.\, For a more detailed insight into this problem, a scaling solution has been proposed \cite{Mueller:2006,Mueller:2006II} for following the evolution between the instabilities taking place in the initial phase and the system in the final equilibration. This solution, depending on one single parameter $\delta$, matches onto the intermediate stage and/or the late stage of the evolution of the system given by the bottom-up thermalization, where the $\delta$ accepts values in the range of \,$0 <\delta < 10/21$.\, As in the case of the Glasma, here the $\delta$ again describes the asymmetry between the transverse and longitudinal scales of the gluon interactions. The scaling solution makes the bottom-up scenario getting modified to the more general {\it m}'bottom-up scenario. The number $10/21$ is the absolute limit the $\delta$ may accept in the {\it m}'bottom-up. So the proposed scaling solution is given by the following set of equations:
\begin{eqnarray}
& & N_{s} \sim {Q_{s}^{3} \over \alpha_{s} (Q_{s}\tau)^{4/3 - \delta}}\,,\,\,\,\,\,\,\,\,\,\,\,\,\,k_{s} \sim {Q_{s} \over
(Q_{s}\tau)^{1/3 - 2\delta/5}}\,,
\nonumber\\
& & \alpha_{s}f_{s} \sim {1 \over (Q_{s}\tau)^{1/3 + \delta/5}}\,,\,\,\,\,\,\,\,\,\,\,M_{D} \sim {Q_{s} \over
(Q_{s}\tau)^{1/2 - 3\delta/10}}\,,
\label{eqn_scaling}
\end{eqnarray}
where the $N_{s}$ is the number density of the soft gluons, the $k_{s}$ is the soft gluon momentum, the $f_{s}$ is the soft gluon occupation number, and the $M_{D}$ is the Debye mass.

\subsection{The two momentum scales of the {\it m}'bottom-up scenario}

There might be a question about whether or not similar (or different) momentum scales, such as the discussed Glasma $\Lambda$ and $\Lambda_{s}$ scales, are applicable for the {\it m}'bottom-up. Conceptually it should take place, since as it was stated before the main qualitative features of the solution of the Boltzmann equation can be described with an assumption that the evolution is dominated by such scales. In the case of the {\it m}'bottom-up let us designate these scales as $\Omega$ and $\Omega_{s}$, and check whether or not such scales can really exist. Making use of the second formula of Eq.\,(\ref{eqn_Glasma_distributions}), we assume that an equivalent relation (with \,$p \approx \omega_{p}$\, at \,$M_{D} \ll \omega_{p}$) can exist in the {\it m}'bottom-up picture, though it is not the case in the original bottom-up:
\beq
f(p) \sim {1 \over \alpha_{s}}{\Omega_{s} \over p}\,\,\,\,\,\mbox{at}\,\,\,\Omega_{s} < p < \Omega\,.
\label{bottom_distribution_g}
\eeq
These scales are related to the soft gluon density $N_{s}$:
\begin{equation}
N_{s} = \int^{\Omega} d^{3}p f(p) \sim \int^{\Omega} p^{2}dp\,{1 \over \alpha_{s}}{\Omega_{s} \over p} \sim {1 \over \alpha_{s}} \Omega^{2} \Omega_{s}\,,
\label{eqn_bottom_scales1}
\end{equation}
and to the Debye mass $M_{D}$:
\begin{equation}
M_{D}^{2} = \alpha_{s}\int^{\Omega} d^{3}p {f(p) \over p} \sim \alpha_{s}\int^{\Omega} p^{2}dp\,{1 \over \alpha_{s}}{\Omega_{s} \over p^{2}} \sim \Omega\Omega_{s}\,.
\label{eqn_bottom_scales2}
\end{equation}
For the quark occupation number we use \,$f_{q} \sim 1$,\, such that
\begin{equation}
N_{q} = \int^{\Omega} d^{3}p f(p) \sim \Omega^{3}\,.
\label{eqn_bottom_scales3}
\end{equation}
As in the case of Glasma, this also comes from the reasoning that at the earliest times we have \,$N_{q} \sim \alpha_{s} N_{h} \ll N_{h}$\, but at later times (for example, at thermalization) \,$N_{q} \sim N_{s}$.\, Here, the $N_{h}$ is the number density of the primary hard gluons (initial gluons).

So how can we make sure that the scales $\Omega$ and $\Omega_{s}$ really hold in the {\it m}'bottom-up scenario ? One may see this if we look back at Eq.\,(\ref{eqn_Glasmagluon}) and Eq.\,(\ref{eqn_GlasmaDebye}), and start with the analogical equations for the {\it m}'bottom-up such as obtained in Eq.\,(\ref{eqn_bottom_scales1}) and Eq.\,(\ref{eqn_bottom_scales2}):
\begin{equation}
N_{s} \sim {1 \over \alpha_{s}}\Omega^{2} \Omega_{s}\,,
\label{eqn_bottomgluon}
\end{equation}
\begin{equation}
M_{D}^{2} \sim \Omega\Omega_{s}\,,
\label{eqn_bottomDebye}
\end{equation}
with the following requirements:
\begin{equation}
\Omega(\tau_{0}) = \Omega_{s}(\tau_{0}) \sim Q_{s}\,,
\label{eqn_bottom_initial}
\end{equation}
and
\begin{equation}
\Omega(\tau_{therm}) \sim T_{therm} \approx T_{in,QGP}\,,\,\,\,\,\Omega_{s}(\tau_{therm}) \sim \alpha_{s} \Omega(\tau_{therm})\,.
\label{eqn_bottom_boundary}
\end{equation}

Thereby, using Eqs.\,(\ref{eqn_bottomgluon}), (\ref{eqn_bottomDebye}), (\ref{eqn_bottom_initial}) and (\ref{eqn_bottom_boundary}) at the initial time \,$\tau_{0} \sim Q_{s}^{-1}$,\, we shall have the ``initial conditions":
\begin{equation}
N_{h} \sim { Q_{s}^{3} \over \alpha_{s}(Q_{s}\tau_{0})}\,,
\label{eqn_bottomgluon_in}
\end{equation}
\begin{equation}
M_{D}^{2} \sim {Q_{s}^{2} \over (Q_{s}\tau_{0})}\,,
\label{eqn_bottomDebye_in}
\end{equation}
as it holds in the original bottom-up ansatz as well. At the thermalization time there should be the following conditions, which we name as ``boundary conditions":
\begin{equation}
N_{s} \sim T_{therm}^{3}\,,
\label{eqn_bottomgluon_fin}
\end{equation}
\begin{equation}
M_{D}^{2} \sim \alpha_{s} T_{therm}^{2}\,.
\label{eqn_bottomDebye_fin}
\end{equation}

\subsection{The derivation of the soft gluon number density, Debye mass, energy density and entropy density via thermalization temperature}

In this section we show how Eqs.\,(\ref{eqn_bottomgluon_fin}) and (\ref{eqn_bottomDebye_fin}) can be derived in the {\it m}'bottom-up. For this purpose we should use the $N_{s}$ and $M_{D}$ from Eq.\,(\ref{eqn_scaling}), along with the thermal bath temperature $T(\tau)$ from \cite{Mueller:2006,Mueller:2006II}:
\begin{equation}
N_{s} \sim {Q_{s}^{3} \over \alpha_{s} (Q_{s}\tau)^{4/3 - \delta}}\,,
\label{eqn_bottomNs}
\end{equation}
\begin{equation}
M_{D}^{2} \sim {Q_{s}^{2} \over (Q_{s}\tau)^{1 - 3\delta/5}}\,,
\label{eqn_bottomMD}
\end{equation}
\begin{equation}
T^{2} \sim Q_{s}^{2}\,\alpha_{s}^{2(35 - 78\delta)/(39\delta - 10)}(Q_{s}\tau)^{2(15 - 36\delta)/(39\delta - 10)}\,.
\label{eqn_bottom_temp}
\end{equation}
Independent of what value the $\delta$ takes, as long as \,$\delta > 1/3$,\, the temperature in Eq.\,(\ref{eqn_bottom_temp}) and the scaling solution in Eq.\,(\ref{eqn_scaling}) match up onto the final stage of the original bottom-up only at the final time \,$Q_{s}\tau \sim \alpha_{s}^{-13/5}$.\, In this case the temperature in Eq.\,(\ref{eqn_bottom_temp}) reduces to \,$T \sim Q_{s} \alpha_{s}^{2/5}$\, which is independent of $\delta$. 

\begin{figure}[h]
\begin{center}
\vspace{0.2cm}
\includegraphics[width=10cm]{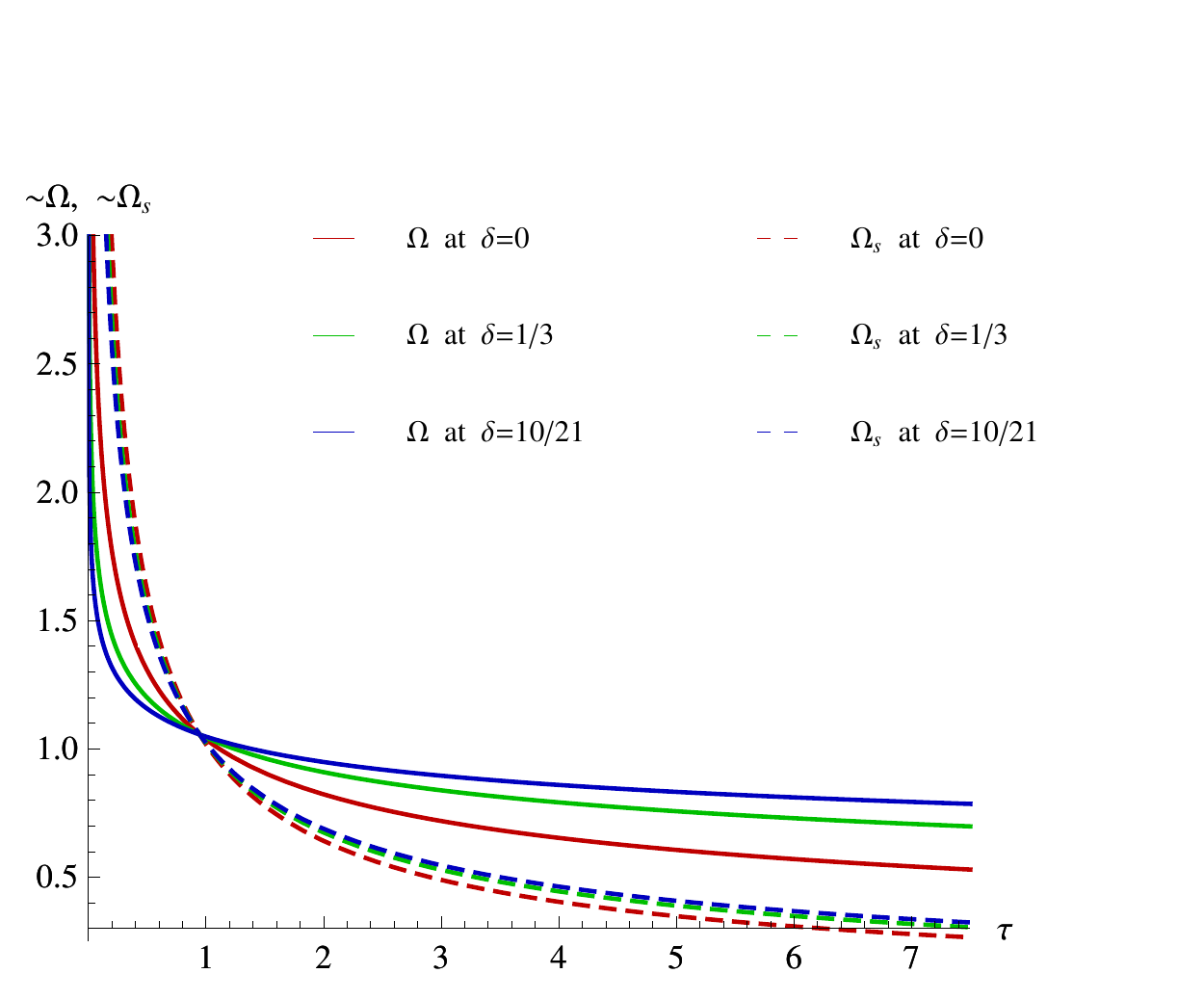}
\caption{The two scales $\Omega$ and $\Omega_{s}$ as functions of the proper time $\tau$, at three values of $\delta$ shown from top to bottom as \,$\delta=0$,\, \,$\delta=1/3$\, and \,$\delta=10/21$.}
\label{OmegaOmegas}
\end{center} 
\end{figure}

Consequently, we should search for some solutions for $N_{s}$ and $M_{D}$ that should be $\delta$-independent. We start with Eq.\,(\ref{eqn_bottomMD}) and Eq.\,(\ref{eqn_bottomDebye}) such as
\begin{equation}
M_{D}^{2} \sim{Q_{s}^{2} \over (Q_{s}\tau)^{1- 3\delta/5}}\,\,\,\,\,\,\,\,\,\,\,\,\mbox{and}\,\,\,\,\,\,\,\,\,\,\,\,M_{D}^{2} \sim \Omega \Omega_{s}\,,
\label{eqn_eqns1}
\end{equation}
and solve them for $\Omega_{s}$:
\begin{equation}
\Omega_{s} \sim {1 \over \Omega} {Q_{s}^2 \over (Q_{s}\tau)^{1- 3\delta/5}}\,.
\label{eqn_Omegas}
\end{equation}
But still the scale $\Omega$ (dependent on $\tau$) stands in the denominator of  Eq.\,(\ref{eqn_Omegas}). Then making use of  Eqs.\,(\ref{eqn_bottomNs}) and (\ref{eqn_bottomgluon}) 
\begin{equation}
N_{s} \sim {Q_{s}^{3} \over \alpha_{s} (Q_{s}\tau)^{4/3 - \delta}}\,\,\,\,\,\,\,\,\,\,\,\,\mbox{and}\,\,\,\,\,\,\,\,\,\,\,\,N_{s} \sim {1 \over \alpha_{s}} \Omega^{2} \Omega_{s}\,,
\label{eqn_eqns2}
\end{equation}
along with (\ref{eqn_Omegas}) and solving them for $\Omega$, we further find the scale $\Omega$ as a function of $\tau$:
\begin{equation}
\Omega \sim Q_{s} {\Lb 1 \over Q_{s}\tau \Rb}^{(5 - 6\delta)/15}\,.
\label{eqn_Omegatau}
\end{equation}
Thereby, by inserting the $\Omega$ from Eq.\,(\ref{eqn_Omegatau}) into the denominator of Eq.\,(\ref{eqn_Omegas}), we obtain the scale $\Omega_{s}$ as a function of $\tau$.
\begin{equation}
\Omega_{s} \sim Q_{s} {\Lb 1 \over Q_{s}\tau \Rb}^{(10 - 3\delta)/15}\,.
\label{eqn_Omegastau}
\end{equation}
In Fig.\,\ref{OmegaOmegas}, the derived time-dependent scales $\Omega$ and $\Omega_{s}$ are parametrically depicted at selected three values of the $\delta$ parameter.

With these two derivations of $\Omega$ and $\Omega_{s}$, we can parametrically find the thermalization time in the {\it m}'bottom-up scenario. We can get it using the condition in Eq.\,(\ref{eqn_bottom_boundary}):
\begin{equation}
Q_{s} {\Lb 1 \over Q_{s}\tau_{therm} \Rb}^{(10 - 3\delta)/15} \sim \alpha_{s} Q_{s} {\Lb 1 \over Q_{s}\tau_{therm} \Rb}^{(5 - 6\delta)/15}\,,
\end{equation}
which ultimately gives
\begin{equation}
Q_{s}\tau_{therm} \sim \alpha_{s}^{-{15/(5 + 3\delta)}}\,.
\label{eqn_bottom_therm}
\end{equation}

As a next step we divide both sides of Eq.\,(\ref{eqn_bottomNs}) and Eq.\,(\ref{eqn_bottom_temp}) (with $T^{3}$) on each other, and both sides of Eq.\,(\ref{eqn_bottomMD}) and Eq.\,(\ref{eqn_bottom_temp}) again on each other (also by using \,$Q_{s}\tau_{therm}$\, from Eq.\,(\ref{eqn_bottom_therm})). Then we shall have the following two formulas:
\bea
N_{s} & \sim & \alpha_{s}^{-\left[ 1 + 3(35 - 78\delta)/(39\delta - 10) \right]} \left( \alpha_{s}^{-15/(5 + 3\delta)} \right)^{-\left[ 3(15 - 36\delta)/(39\delta - 10) + (4 - 3\delta)/3 \right]} T_{therm}^{3}
\nonumber\\
& \sim & \alpha_{s}^{[-195\delta + 50 - 117\delta^{2} + 30\delta - 525 - 315\delta + 1170\delta + 702\delta^{2}]/[(5 + 3\delta)(39\delta - 10)]} \times
\nonumber\\
&  & \times\,\,\alpha_{s}^{[675 - 1620\delta + 780\delta - 200 - 585\delta^{2} + 150\delta]/[(5 + 3\delta)(39\delta - 10)]}  T_{therm}^{3} 
\nonumber\\
& \sim & \alpha_{s}^{[585\delta^{2} - 585\delta^{2} + 690\delta - 690\delta - 475 + 475]/[(5 + 3\delta)(39\delta - 10)]}  T_{therm}^{3}\,,
\label{eqn_bottomNsright}
\eea
which reduces to \,$N_{s} \sim T_{therm}^{3}$,\, that is to say the same as Eq.\,(\ref{eqn_bottomgluon_fin}). And
\bea
M_{D}^{2} & \sim & \alpha_{s}^{ -2(35 - 78\delta)/(39\delta - 10)} \left( \alpha_{s}^{-15/(5 + 3\delta)} \right)^{ - \left[ 1 - 3\delta/5 + 2(15 - 36\delta)/(39\delta - 10) \right]} T_{therm}^{2}
\nonumber\\
& \sim & \alpha_{s}^{[-350 -210\delta + 780\delta + 468\delta^{2} + 585\delta - 150 - 351\delta^{2} + 90\delta + 450 - 1080\delta]/[(5 + 3\delta)(39\delta - 10)]} T_{therm}^{2}
\nonumber\\
& \sim & \alpha_{s}^{[117\delta^{2} + 165\delta - 50]/[(5 + 3\delta)(39\delta - 10)]} T_{therm}^{2}\,,
\label{eqn_bottomMDright}
\eea
which reduces to \,$M_{D}^{2} \sim \alpha_{s} T_{therm}^{2}$,\, the same as Eq.\,(\ref{eqn_bottomDebye_fin}). Alternatively, an analogous outcome can be demonstrated if we use the $\Omega$ and $\Omega_{s}$ scales.
\bea
\!\!\!\!\!\!\!\!\!\!
\frac{N_{s}}{T^{3}}|_{therm} & \sim & \Lb {1 \over \alpha_{s}} \Omega^{2} \Omega_{s} \Rb / \Lb Q_{s}^{3}\,\alpha_{s}^{3(35 - 78\delta)/(39\delta - 10)}\,(Q_{s}\tau)^{3(15 - 36\delta)/(39\delta - 10)} \Rb|_{therm} \Rightarrow
\nonumber\\ 
N_{s} & \sim & T_{therm}^{3}\,,
\label{eqn_NsT}
\eea
and
\bea
\!\!\!\!\!\!\!\!\!\!
\frac{M_{D}^{2}}{T^{2}}|_{therm} & \sim & \Lb \Omega\Omega_{s} \Rb / \Lb Q_{s}^{2}\,\alpha_{s}^{2(35 - 78\delta)/(39\delta - 10)}\,(Q_{s}\tau)^{2(15 - 36\delta)/(39\delta - 10)} \Rb|_{therm} \Rightarrow
\nonumber\\ 
M_{D}^{2} & \sim & \alpha_{s} T_{therm}^{2}\,,
\label{eqn_MDT}
\eea
Consequently, the results in Eqs.\,(\ref{eqn_bottomNsright}) and (\ref{eqn_bottomMDright}) are independent of any value the $\delta$ accepts. Note that the thermalization temperature $T_{therm}$ of the {\it m}'bottom-up parton matter is assumed to be parametrically of the same order as the initial temperature of the Quark-Gluon Plasma $T_{in,QGP}$.

However, this is not the end of the story. An equilibrated system of soft gluons at the thermalization temperature $T_{therm}$ must satisfy the condition \,$\epsilon_{s} \sim T_{therm}^{4}$\, for the energy density. Needless to say that we should also be able to prove the validity of this relation in our ansatz. From \cite{Mueller:2006,Mueller:2006II} we find the energy density of the soft sector expressed via the soft gluon number density and momentum, \,$\epsilon_{s} \sim N_{s}k_{s}$.\, Then from the scaling solution of Eq.\,(\ref{eqn_scaling}) we will have the following expression:
\begin{equation}
\epsilon_{s} \sim {Q_{s}^{4} \over \alpha_{s} (Q_{s}\tau)^{(25 - 21\delta)/15}}\,,
\label{eqn_en_density1}
\end{equation}
By dividing both sides of Eq.\,(\ref{eqn_en_density1}) and Eq.\,(\ref{eqn_bottom_temp}) (with $T^{4}$) on each other along with using \,$Q_{s}\tau_{therm}$\, from Eq.\,(\ref{eqn_bottom_therm}), the energy density becomes
\bea
\epsilon_{s} & \sim & \alpha_{s}^{-\left[ 1 + 4(35 - 78\delta)/(39\delta - 10) \right]} \left( \alpha_{s}^{-15/(5 + 3\delta)} \right)^{-\left[ (25 - 21\delta)/15 + 4(15 - 36\delta)/(39\delta - 10) \right]} T_{therm}^{4}
\nonumber\\
& \sim & \alpha_{s}^{[-195\delta + 50 - 117\delta^{2} + 30\delta - 700 - 420\delta + 1560\delta + 936\delta^{2}]/[(5 + 3\delta)(39\delta - 10)]} \times
\nonumber\\
&  & \times\,\,\alpha_{s}^{[975\delta- 250 - 819\delta^{2} + 210\delta + 900 - 2160\delta]/[(5 + 3\delta)(39\delta - 10)]} T_{therm}^{4} 
\nonumber\\
& \sim & \alpha_{s}^{[0 + 0 + 0]/[(5 + 3\delta)(39\delta - 10)]} T_{therm}^{4}\,.
\label{eqn_en_density2}
\eea

In addition to the above derivations, one can also pay attention to the following observation. Let us assume for a moment that 
the scale $\Omega$ from Eq.\,(\ref{eqn_Omegatau}) and the scale \,$\Lambda \sim Q_{s} \Lb 1/Q_{s}\tau \Rb^{(1 + 2\delta^{\prime})/7}$\, from \cite{Blaizot:2011xf} are parametrically the same. 
\begin{equation}
\Omega \sim \Lambda\,\,\,\,\,\,\,\,\,\,\Rightarrow\,\,\,\,\,\,\,\,\,\,Q_{s} {\Lb 1 \over Q_{s}\tau \Rb}^{(5 - 6\delta)/15} \sim Q_{s}{\Lb 1 \over Q_{s}\tau \Rb}^{(1 + 2\delta^{\prime})/7}\,.
\label{eqn_OmegaLambda}
\end{equation}
Mathematically this assumption is valid if 
\begin{equation}
\delta = \frac{10 - 15\delta^{\prime}}{21}\,,
\label{eqn_delta_deltaprime}
\end{equation}
or vice versa
\begin{equation}
\delta^{\prime} = \frac{10 - 21\delta}{15}\,.
\label{eqn_deltaprime_delta}
\end{equation}
Then what will happen if this $\delta \leftrightarrow \delta^{\prime}$ relation is being inserted into the r.h.s. of Eq.\,(\ref{eqn_Omegastau}) ? This results in what follows:
\begin{equation}
\Omega_{s} \sim Q_{s} {\Lb 1 \over Q_{s}\tau \Rb}^{(10 - 3\delta)/15}\,\,\,\,\,\rightarrow\,\,\,\,\,Q_{s} {\Lb 1 \over Q_{s}\tau \Rb}^{(4 + \delta^{\prime})/7}\,.
\label{eqn_OmegasLambda}
\end{equation}
But the r.h.s of Eq.\,(\ref{eqn_OmegasLambda}) is actually the scale $\Lambda_{s}$ from \cite{Blaizot:2011xf}, which is \,$\Lambda_{s} \sim Q_{s} \Lb 1/Q_{s}\tau \Rb^{(4 + \delta^{\prime})/7}$.\, In this connection let us again take a look at Ref.\,\cite{Blaizot:2011xf} but for the gluon density and Debye mass
\begin{equation}
N_{g} \sim {Q_{s}^{3} \over \alpha_{s} (Q_{s}\tau)^{(6 + 5\delta^{\prime})/7}}\,\,\,\,\,\,\,\,\,\,\,\,\,\,\,\,\,\,\,\,M_{D}^{2} \sim {Q_{s}^{2} \over (Q_{s}\tau)^{(5 + 3\delta^{\prime})/7}}\,,
\label{eqn_Glasma_NgMD}
\end{equation}
as well as for the thermalization time, which is obtained with the condition of Eq.\,(\ref{eqn_Glasma_scales})
\begin{equation}
Q_{s}\tau_{therm} \sim \alpha_{s}^{-{7/(3 - \delta^{\prime})}}\,.
\label{eqn_Glasma_tautherm}
\end{equation}
Besides, we also wish to take a look at the energy density in gluon modes  
\begin{equation}
\epsilon_{g} \sim \frac{1}{\alpha_{s}}\Lambda_{s}\Lambda^{3}\,,
\label{eqn_Glasma_en_density1}
\end{equation}
which, in the case of the longitudinal expansion under the assumption of the parameter $\delta^{\prime}$ independent of time, is represented as an equation for the evolution of the energy density:
\begin{equation}
\epsilon_{g}(\tau) \sim \epsilon_{g}(\tau_{0}) \Lb \frac{1}{Q_{s}\tau} \Rb^{1 + \delta^{\prime}}\,,
\label{eqn_Glasma_en_density2}
\end{equation}
Thereby, by making use of Eq.\,(\ref{eqn_deltaprime_delta}) it is easy to check that the following transformations take place:
\begin{equation}
N_{g}\,\,\mbox{of\,\,Eq.\,(\ref{eqn_Glasma_NgMD})}\,\,\,\,\,\rightarrow\,\,\,\,\,N_{s}\,\,\mbox{of\,\,Eq.\,(\ref{eqn_scaling})}\,;
\label{eqn_NgNs}
\end{equation}
\begin{equation}
M_{D}\,\,\mbox{of\,\,Eq.\,(\ref{eqn_Glasma_NgMD})}\,\,\,\,\,\rightarrow\,\,\,\,\,M_{D}\,\,\mbox{of\,\,Eq.\,(\ref{eqn_scaling})}\,;
\label{eqn_MDMD}
\end{equation}
\begin{equation}
\tau_{therm}\,\,\mbox{of\,\,Eq.\,(\ref{eqn_Glasma_tautherm})}\,\,\,\,\,\rightarrow\,\,\,\,\,\tau_{therm}\,\,\mbox{of\,\,Eq.\,(\ref{eqn_bottom_therm})}\,;
\label{eqn_MDMD}
\end{equation}
\begin{equation}
\epsilon_{g}\,\,\mbox{of\,\,Eq.\,(\ref{eqn_Glasma_en_density1})}\,\,\,\,\,\rightarrow\,\,\,\,\,\epsilon_{s}\,\,\mbox{of\,\,Eq.\,(\ref{eqn_en_density1})}\,.
\label{eqn_epsilong_epsilons}
\end{equation}
However, these transformations can be valid in the range of \,$10/21 > \delta > 5/21$\, corresponding to the range of \,$0 < \delta^{\prime} < 1/3$,\, This can be seen as follows:
\begin{quote}
at \,$\delta^{\prime} = 0$\, \,$\rightarrow$\, \,$\delta = 10/21$,\,\,\,\,
at \,$\delta^{\prime} = 1/3$\, \,$\rightarrow$\, \,$\delta = 5/21$,
at \,$\delta^{\prime} = 2/3$\, \,$\rightarrow$\, \,$\delta = 0$.
\end{quote}
On the other hand, if formally one needs to recover the static case by setting \,$\delta^{\prime} = -1$,\, corresponding to constant energy density, \,$\epsilon_{g}(\tau) \rightarrow \epsilon_{g}(\tau_{0})$,\, then it will also be the case for the {\it m}'bottom-up scenario, \,$\epsilon_{s}(\tau) \rightarrow \epsilon_{s}(\tau_{0}) \sim Q_{s}^{4}/\alpha_{s}$,\, because in that case \,$\delta =25/21$.

For the energy density one may simultaneously see 
\begin{equation}
\epsilon_{s} \sim \frac{1}{\alpha_{s}}\Omega_{s}\Omega^{3}\,\,\,\,\,\rightarrow\,\,\,\,\,{Q_{s}^{4} \over \alpha_{s} (Q_{s}\tau)^{(25 - 21\delta)/15}}\,.
\label{eqn_en_densities}
\end{equation}
As a last step we wish to prove that the condition for the entropy density in the thermal gluon bath at thermalization is derived in the form of \,$s \sim T_{therm}^{3}$\,. 
\bea
\!\!\!\!\!\!\!\!\!\!
\frac{s}{T^{3}}|_{therm} & \sim & \Omega^{3} / \Lb Q_{s}^{3}\,\alpha_{s}^{3(35 - 78\delta)/(39\delta - 10)}(Q_{s}\tau)^{3(15 - 36\delta)/(39\delta - 10)} \Rb|_{therm} \Rightarrow
\nonumber\\ 
s & \sim & \alpha_{s}^{-3(35 - 78\delta)/(39\delta - 10)} \left( \alpha_{s}^{-15/(5 + 3\delta)} \right)^{-\left[ (15 - 18\delta)/15 + 3(15 - 36\delta)/(39\delta - 10) \right]} T_{therm}^{3}
\nonumber\\
& \sim & \alpha_{s}^{[-525 - 315\delta + 1170\delta + 702\delta^{2} + 585\delta - 150 - 702\delta^{2} + 180\delta + 675 -1620\delta]/[(5 + 3\delta)(39\delta - 10)]} T_{therm}^{3} 
\nonumber\\
& \sim & \alpha_{s}^{[0 + 0 + 0]/[(5 + 3\delta)(39\delta - 10)]} T_{therm}^{3}\,.
\label{eqn_entropy}
\eea

And at the end of this section we mention one of the results from \cite{Mueller:2006}. If \,$\delta > 1/3$,\, the solution in Eq.\,(\ref{eqn_scaling}) changes the character at a time $\tau$ given by
\begin{equation}
Q_{s}\tau \sim (1/\alpha_{s})^{15/(5 + 3\delta)}\,,
\label{eqn_bottom_tau1}
\end{equation}
at which $f_{s} \sim 1$. In that case the scaling solution goes into evolution much like the final phase of the original bottom-up where the soft gluons are thermalized and the harder gluons feed energy into the soft thermalized system causing the temperature to rise with the time until the whole system is thermalized. We see that Eq.\,(\ref{eqn_bottom_therm}), which we get after introducing the scales $\Omega$ and $\Omega_{s}$ into the {\it m}'bottom-up scenario, turns out to be the same as Eq.\,(\ref{eqn_bottom_tau1}), though in our approach the corresponding time is already considered to be as the thermalization time of the system stemming from the thermalization condition of Eq.\,(\ref{eqn_bottom_boundary}).

\section{The thermalization time and the thermalization temperature of the {\it m}'bottom-up parton system at RHIC $\sqrt{s_{NN}} = 200\,GeV$}

As an example of our formalism, one may find values of the thermalization time (from Eq.\,(\ref{eqn_bottom_therm})) and temperature (from Eq.\,(\ref{eqn_bottom_temp})) at RHIC $\sqrt{s_{NN}} = 200\,GeV$ collision energy by following the procedure that has been used in \cite{Baier:2002}. So for the thermalization time we have
\begin{equation}
\tau_{therm} = C_{therm}\,\alpha_{s}^{-15/(5 + 3\delta)} Q_{s}^{-1}\,,
\label{eqn_therm_time}
\end{equation}
where $C_{therm}$ is the so called thermalization constant. For the thermalization temperature we have
\begin{equation}
T_{therm} = C_{T}\,\alpha_{s}^{(35 - 78\delta)/(39\delta - 10)}(Q_{s}\tau_{therm})^{(15 - 36\delta)/(39\delta - 10)}Q_{s}\,,
\label{eqn_therm_temp1}
\end{equation}
which by using Eq.\,(\ref{eqn_therm_time}) reduces to
\begin{equation}
T_{therm} \simeq 0.16543\,C\,C_{therm}^{(15 - 36\delta)/(39\delta - 10)}\,\alpha_{s}^{(5 - 6\delta)/(5 + 3\delta)} Q_{s}\,,
\label{eqn_therm_temp2}
\end{equation}
where we also use the numerical constant $C_{T}$ expressed by the ``gluon liberation'' coefficient:
\begin{equation}
C_{T} \simeq {15 \over 8\pi^{5}}\,N_{c}^{3}\,C \simeq 0.16543\,C\,.
\label{eqn_numerical}
\end{equation}
The $C$ is a parameter that links the number of gluons in the nucleus wave function to the number of gluons which are freed during the collision.

As discussed already, the number of gluons increases with the time $\tau$, because the  hard gluons degrade and the soft ones are formed and start to dominate in the system. Such an increase of the number of gluons must be $2$ or larger, and can be found as the following ratio.
\bea
R & = & {\left[ N_{s}(\tau)(Q_{s}\tau) \right]|_{\tau_{therm}} \over \!\!\!\!\!\left[ N_{h}(\tau)(Q_{s}\tau) \right]|_{\tau_{0}}} \geq 2\,\Rightarrow
\nonumber\\
R & \simeq & 0.13061\,C^{2}\,C_{therm}^{(35 - 69\delta)/(39\delta - 10)}\,\alpha_{s}^{(5 - 15\delta)/(5 + 3\delta)} \geq 2\,\Rightarrow
\nonumber\\
2 & \leq & 0.13061 \left( {4\pi \over 9} \right)^{(5 - 15\delta)/(5 +3\delta)} C^{2}\,C_{therm}^{(35 - 69\delta)/(39\delta - 10)} \left( \ln{\!\left( {Q_{s}^{2} \over \Lambda_{QCD}^{2}} \right)} \right)^{(15\delta - 5)/(5 + 3\delta)}
\label{eqn_ratio}
\eea
where we make use of Eq.\,(\ref{eqn_therm_time}) and Eq.\,(\ref{eqn_therm_temp2}) along with the formulas shown in what follows:
\begin{equation}
N_{s}(\tau_{therm}) = 2(N_{c}^{2} - 1){\zeta(3) \over \pi^{2}}\,T_{therm}^{3}\,,
\label{eqn_nsoft}
\end{equation}
\begin{equation}
N_{h}(\tau_{0}) = C\,{(N_{c}^{2} - 1)Q_{s}^{3} \over 4\pi^{2}N_{c}\alpha_{s}\,(Q_{s}\tau_{0})}\,,
\label{eqn_nhard}
\end{equation}
\begin{equation}
\alpha_{s}(Q_{s}^{2}) \simeq \frac{4\pi}{\left( 11- {2 \over 3}N_{f} \right) \ln{\!\left( {Q_{s}^{2} \over \Lambda_{QCD}^{2}} \right)}}\,.
\label{eqn_alpha}
\end{equation}
For $N_{c} = 3$ we take $N_{f} = 3$. Also, by using the well-known relation at midrapidity
\begin{equation}
Q_{s}^{2}(A,\sqrt{s}) = Q_{0}^{2}(A,\sqrt{s_{0}}) \left( \frac{\sqrt{s}}{\sqrt{s_{0}}} \right)^{\lambda/(1 + \lambda/2)}\,,
\end{equation}
with $\lambda = 0.288$ \cite{Kharzeev:2004}, we shall have the saturation momentum $Q_{s}^{2} \simeq 1.115\,GeV^{2}$ at $\sqrt{s_{NN}} = 200\,GeV$ in Au+Au collisions, which is obtained from $Q_{0}^{2} = 1\,GeV^{2}$ at $\sqrt{s_{NN}} = 130\,GeV$ that has been used in \cite{Baier:2002}. Thus, the ratio $R$ in Eq.\,(\ref{eqn_ratio}) can be $2$ or larger if the overall constant \,$C^{2}\,C_{therm}^{(35 - 69\delta)/(39\delta - 10)}$\, is taken adequately.

We need also one more formula with $C$ and $C_{therm}$ after which these parameters can be determined. That formula is derived by comparing the calculable charged hadron multiplicity at midrapidity at $\sqrt{s_{NN}} = 200\,GeV$ with that from RHIC Au+Au data. As a reference value we use the result by the PHOBOS collaboration \cite{PHOBOS:2002}, namely \,$3.78 \pm 0.25\,(syst)$.\,Thus,
\begin{equation}
\langle {2 \over N_{part}}{dN_{ch} \over d\eta} \rangle |_{exp} = 3.78 \pm 0.25\,(syst)\,.
\end{equation}
The charged hadron multiplicity can be calculated as
\begin{equation}
\langle {2 \over N_{part}}{dN_{ch} \over d\eta} \rangle \simeq {R\,C \over 3}\,\ln{\!\Lb {Q_{s}^{2} \over \Lambda_{QCD}^{2}} \Rb}\,,
\label{eqn_multip1}
\end{equation}
which is further simplified to be
\bea
& & \langle {2 \over N_{part}}{dN_{ch} \over d\eta} \rangle \simeq
\nonumber\\
& \simeq & 0.04354 \left( {4\pi \over 9} \right)^{(5 - 15\delta)/(5 +3\delta)} C^{3}\,C_{therm}^{(35 - 69\delta)/(39\delta - 10)} \left( \ln{\!\left( {Q_{s}^{2} \over \Lambda_{QCD}^{2}} \right)} \right)^{18\delta/(5 + 3\delta)}\,, 
\label{eqn_multip2}
\eea
where we use Eq.\,(\ref{eqn_ratio}) for the ratio $R$. Here the charged hadron multiplicity will be equal to $3.78$ if the overall constant \,$C^{3}\,C_{therm}^{(35 -69\delta)/(39\delta - 10)}$\, is taken adequately.

\begin{figure}[h!]
\begin{center}
\vspace{0.0cm}
\includegraphics[width=8.0cm]{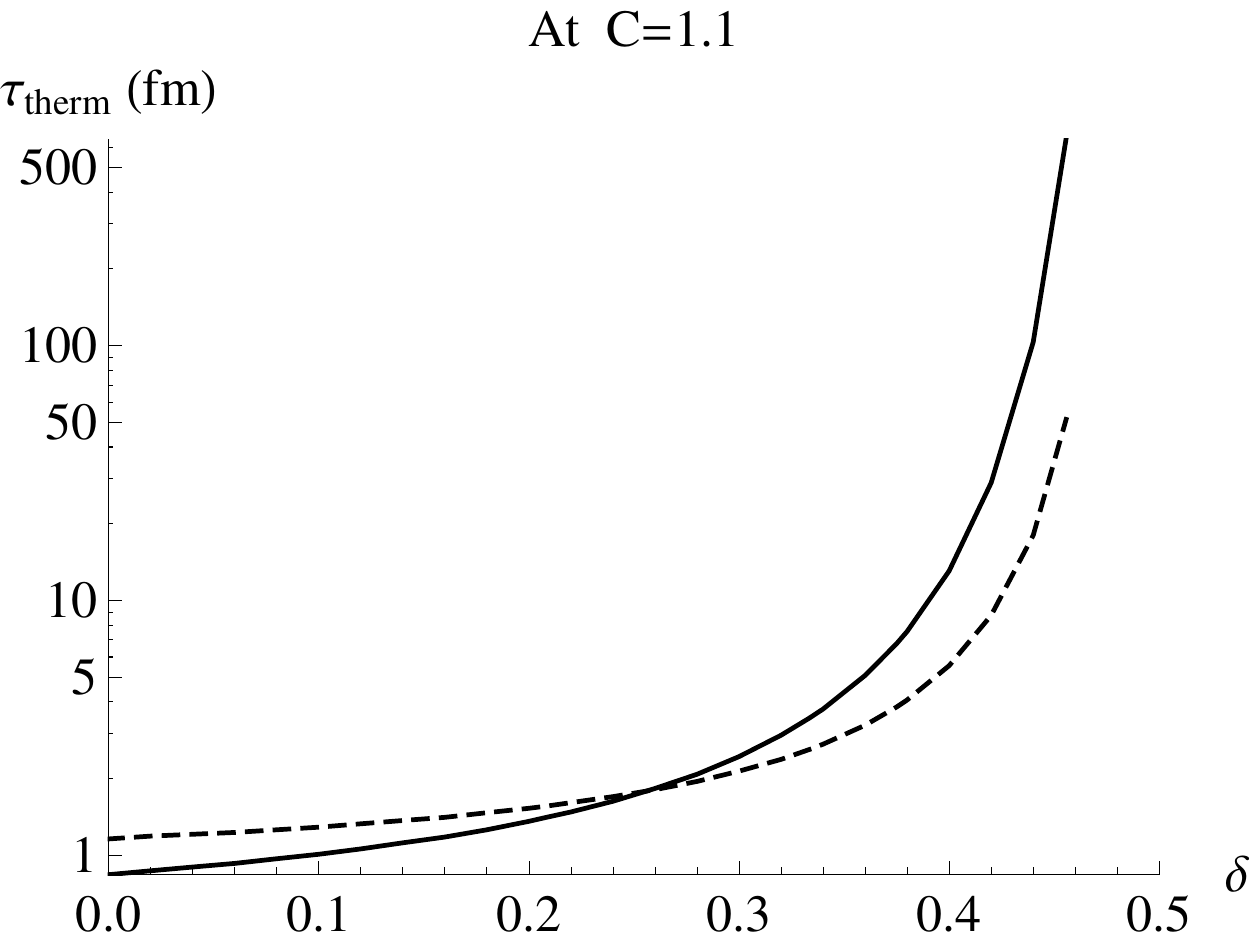}
\caption{The solid and dashed curves show the $\delta$-dependent $\tau_{therm}$ and $\tau_{1}$ at \,$C = 1.1$,\, respectively.}
\label{Omega_tau1}
\end{center} 
\end{figure}

\begin{figure}[h!]
\begin{center}
\vspace{0.0cm}
\includegraphics[width=8.0cm]{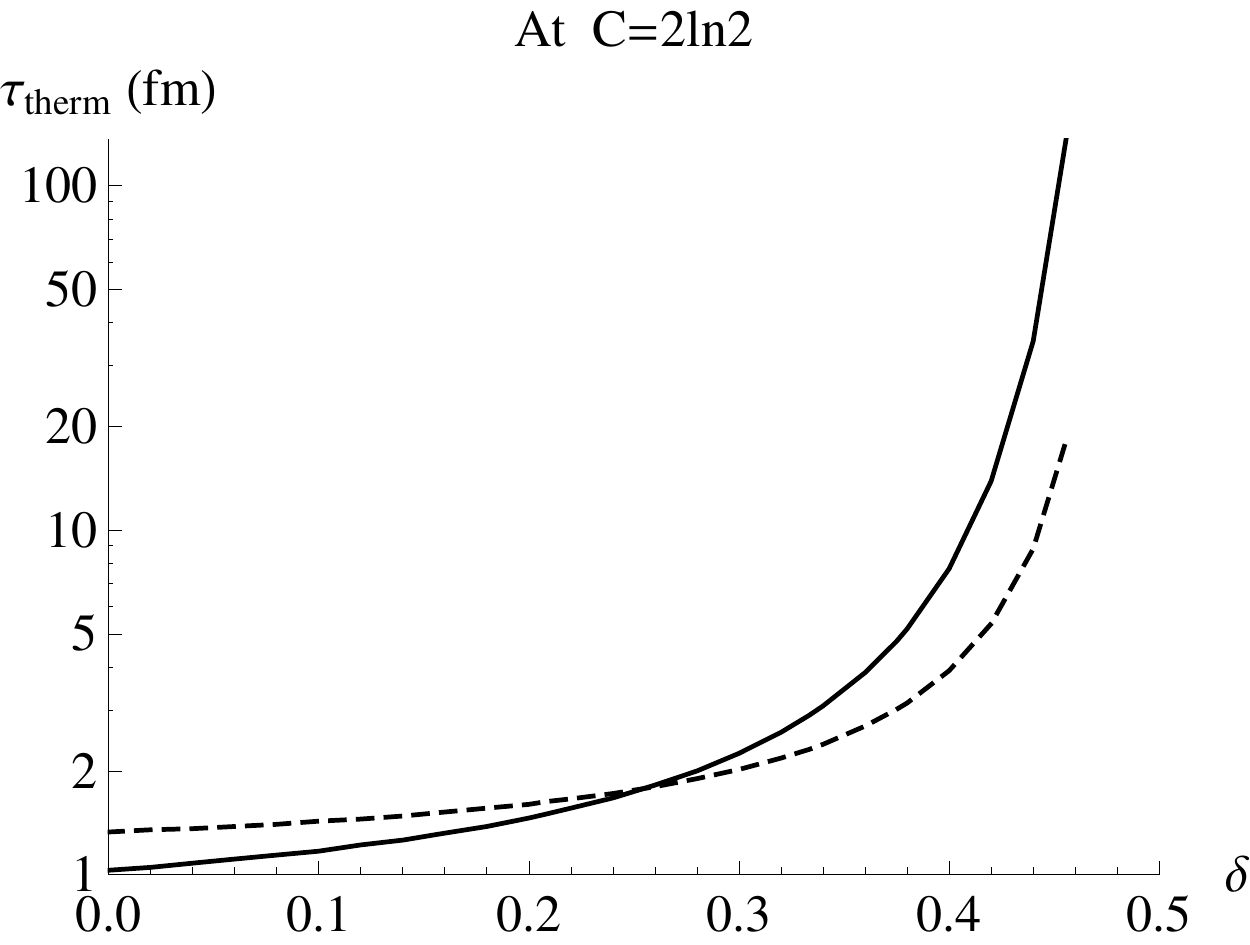}
\caption{The solid and dashed curves show the $\delta$-dependent $\tau_{therm}$ and $\tau_{1}$ at \,$C = 2\ln{2}$,\, respectively.}
\label{Omega_tau2}
\end{center} 
\end{figure}

\begin{figure}[h!]
\begin{center}
\vspace{0.0cm}
\includegraphics[width=8.0cm]{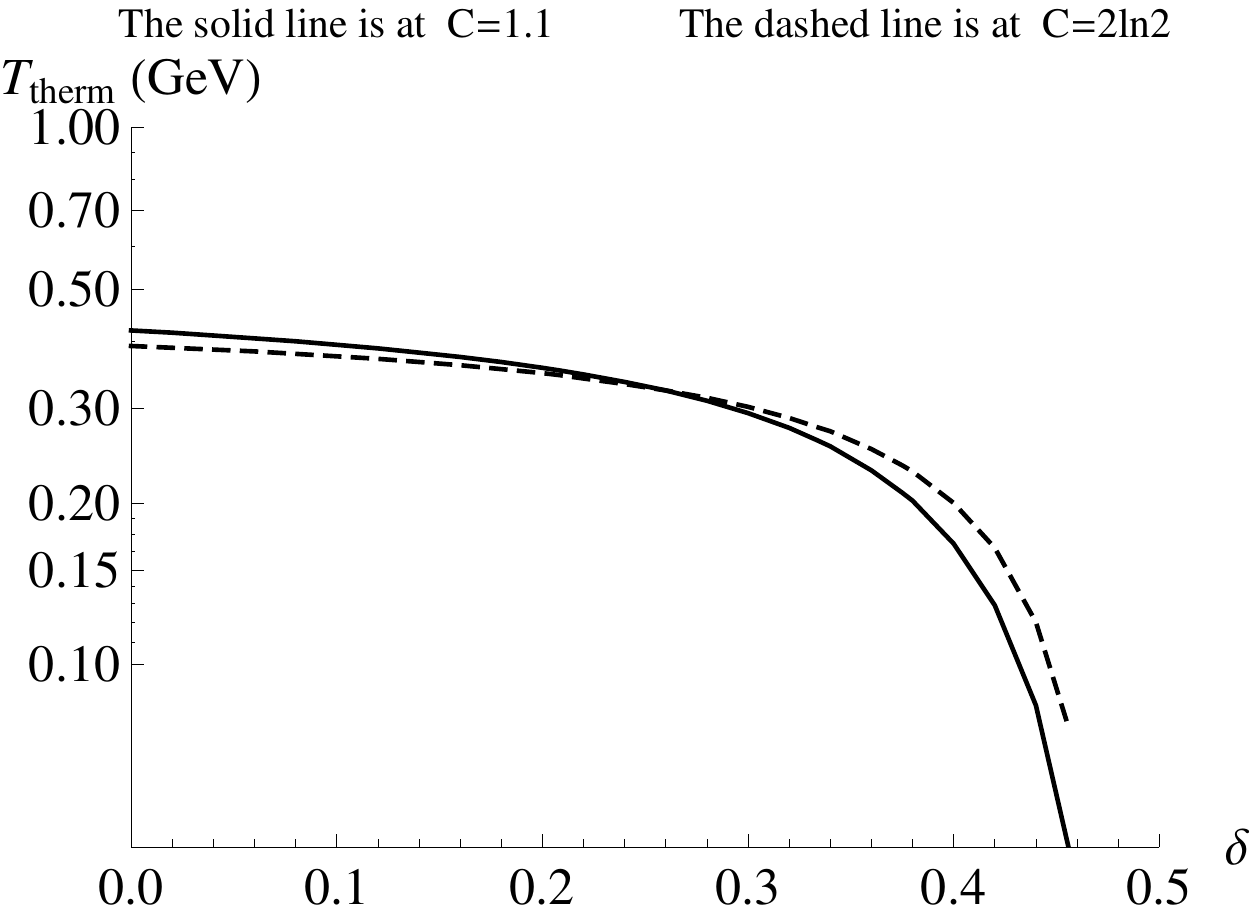}
\caption{The solid and dashed curves show the $\delta$-dependent $T_{therm}$ at \,$C = 1.1$\, and \,$C = 2\ln{2}$,\, respectively.}
\label{Omega_T}
\end{center} 
\end{figure}

Let us now find the $\tau_{therm}$ and $T_{therm}$ as functions of the $\delta$ parameter. We will try two values for the gluon liberation coefficient: the first one calculated in \cite{Kovchegov:2001} with \,$C = 2\ln{2} \simeq 1.386$,\, and the second one in \cite{Lappi:2001} with \,$C = 1.1$. For example, at \,$\delta = 1/3$\, we will have
\begin{displaymath}
C_{therm} \simeq {2.260 \over C^{3/4}}\,\,\, \mbox{and} \,\,\,C_{therm} \geq {1.978 \over C^{2/4}}\,\Rightarrow
\end{displaymath}
\begin{equation}
\Rightarrow\,C \leq 1.704\,\,\,\mbox{and}\,\,\,C_{therm} \geq 1.515\,.
\label{eqn_delta1}
\end{equation}
If we use the above values of the ``gluon liberation" coefficient, then from Eq.\,(\ref{eqn_therm_time}) and Eq.\,(\ref{eqn_therm_temp2}) we shall have the following fixed values for the thermalization time and temperature: 
\bea
\!\!\!\!\!\!\!\mbox{At}\,\,\,\,C = 1.1 \rightarrow\,\,\,\,\,\,\tau_{therm} \simeq 3.45\,fm\,,\,\,T_{therm} \simeq 262\,MeV.
\nonumber \\
\!\!\!\!\!\!\!\mbox{At}\,\,\,\,C = 2\ln{2} \rightarrow \tau_{therm} \simeq 2.90\,fm\,,\,\,T_{therm} \simeq 277\,MeV.
\eea
Also, at \,$C = 1.1 \,\Rightarrow\, R \simeq 3.1$,\, and at \,$C = 2\ln{2}\,\Rightarrow\, R \simeq 2.46$. In Fig.\,\ref{Omega_tau1} and Fig.\,\ref{Omega_tau2} the solid lines show the $\delta$-dependent thermalization time $\tau_{therm}$ at both values of $C$. The $\delta$-dependent thermalization temperature $T_{therm}$ at both values of $C$ is shown in Fig.\,\ref{Omega_T}.

One may put some constraints on the possible values of the parameter $\delta$. It can be done if we find a time $\tau_{1}$ at which $N_{h} = N_{s}$. Namely, by using Eq.\,(\ref{eqn_nhard}) and Eq.\,(\ref{eqn_nsoft}) with Eq.\,(\ref{eqn_therm_temp2}) we derive
\begin{equation}
Q_{s}\tau_{1} = \Lb \frac{1}{0.1306\,C^{2}} \Rb^{(39\delta - 10)/(35 - 69\delta)} \alpha_{s}^{-(95 - 195\delta)/(35 - 69\delta)}\,.
\label{eqn_numerical2}
\end{equation}
It is obvious that at \,$\delta = 1/3$,\, this formula becomes parametrically the same as that of the original bottom-up, i.e., $N_{h} = N_{s}$ at \,$Q_{s}\tau_{1} \sim \alpha_{s}^{-5/2}$.\, Then if we calculate the time $\tau_{1}$ at \,$C = 1.1$\, and \,$C = 2\ln{2}$\, for \,$\delta = 1/3$,\, we shall obtain the following result:
\bea
& & \mbox{At}\,\,\,\,\,C = 1.1\,\,\,\,\,\rightarrow\,\,\,\,\,\tau_{1} \simeq 2.60\,fm, 
\nonumber \\
& & \mbox{At}\,\,\,\,\,C = 2\ln{2}\,\,\,\,\,\rightarrow\,\,\,\,\,\tau_{1} \simeq 2.31\,fm.
\label{eqn_delta2}
\eea
In Fig.\,\ref{Omega_tau1} and Fig.\,\ref{Omega_tau2} the dashed lines show the $\delta$-dependent $\tau_{1}$ at both values of $C$. So since the soft gluons start to overwhelm, in terms of number, the primary hard gluons at \,$\tau_{1} \sim \alpha_{s}^{-(95 - 195\delta)/(35 - 69\delta)}Q_{s}^{-1}$,\, it means that the following inequality must always take place: \,$\tau_{therm} > \tau_{1}$.\, By comparing the results shown in Fig.\,\ref{Omega_tau1} and Fig.\,\ref{Omega_tau2}, we see that the cases of \,$\delta \lesssim 0.26$\, should be excluded. On the other hand, we notice that the evolutionary picture looks highly unlikely at larger values of $\delta$ close to \,$\delta = 10/21$,\, which shows that the system never gets thermalized. Thus, we find that the thermalization time and temperature at values of $\delta$ from a range around \,$\delta = 1/3$ is comparable to those from \cite{Baier:2002}.

However, we should note that the results in Figs.\,\ref{Omega_tau1}, \ref{Omega_tau2} and \ref{Omega_T} are approximate because we make use of the parameter $C_{T}$ from Eq.\,(\ref{eqn_numerical}) (by having it from \cite{Baier:2001,Baier:2002}) in our derivations of this section. It is possible that the $C_{T}$ can be $\delta$-dependent, which perhaps will (or will not) make the curves less steeper at large values of $\delta$, and/or can alter the results for $\tau_{therm}$ and $T_{therm}$ to be similar to those from Ref.\,\cite{Baier:2002} at small values of $\delta$. But the derivation of the $\delta$-dependent $C_{T}$ parameter is beyond the scope of this paper, which however will be considered in another work.

\newpage

{\leftline{\bf Acknowledgements} We are grateful to Larry McLerran and Al Mueller for extremely useful discussions on the subject matter of the paper. We also appreciate the valuable comments made by Alan Dion, Ali Hanks, Richard Petti, Deepali Sharma and Serpil Yalcin. The research of V. Khachatryan and T. K. Hemmick is supported under DOE Contract No. DE-FG02-96-ER40988. The research of M. Chiu is supported under DOE Contract No. DE-AC02-98-CH10886.}

\end{document}